# The dynamics of sex ratio evolution:
## the nature and role of the male subpopulation equilibrium

Krzysztof Argasinski

Institute of Environmental Sciences, Jagiellonian University

Gronostajowa 7, 30-387 Kraków, Poland

e-mail: krzysztof.argasinski@uj.edu.pl, argas1@wp.pl

Tel. +48604535595

Fax: (+48) 12 664 69 12

**CAUTION!!!!!**

**This is a working version of the paper and soon will be replaced by new, extended version containing additional numerical simulations and new results. Numerical simulations revealed some new surprising patterns. In effect, some interpretations of the analytical results from this version are falsified by new results and will be substantially reformulated in the upcoming version.**

**This work was supported by grant N N304 2764 33 from the Polish Ministry of Science and Higher Education.**

## Abstract

The classical approaches to the modeling of sex ratio evolution can be divided into two classes. The first class contains the static strategic models related to the *Dusing Fisher Shaw Mohler* fitness measure, based on the reproductive value of the offspring of the focal female. The second class contains the population genetic models focused on the dynamics of the allele frequencies. The approaches are not fully compatible because the strategic models disregard the role of the male individuals as the passive carriers of the strategy genes. In the previous two papers in this cycle, a new synthetic model combining the strategic analysis with more rigorous genetics was presented. The new model shows that sex ratio self-regulation is a multistage complex process which can be regarded as an example of multilevel selection. One of the elements of this process is the dynamic equilibrium between male and female gene carriers associated with convergence of the dynamics to the manifold termed the *male subpopulation equilibrium* (MSE). This paper attempts to explain this phenomenon and analyze its properties. We show that the MSE phenomenon affects every stage of sex ratio self-regulation (Lemmas 1-4). The MSE plays a crucial role in synchronizing two levels of selection in the double-level selection process. We show that the MSE condition can be generalized as an interesting synergistic property allowing for the estimation of the primary sex ratio of the entire population according to the state of some arbitrarily chosen subpopulation (Lemma 5). We also show that the classical Dusing Fisher Shaw Mohler fitness measure is a biased approximation of the new approach, but that it produces compatible strategic predictions (Lemma 6).

# 1. Introduction

Sex ratio evolution was the first problem in evolutionary theory that involved mathematical reasoning. This was done by German biologist Carl Dusing [1]. He argued that the fitness of a female using a particular sex ratio strategy can be described by the number of her grandchildren. By application of this fitness measure it can be shown that the average strategy of females equal to 0.5 is evolutionarily stable. However, the population genetic approach focused on tracing the gene frequencies produces different predictions [2,3,4]. It shows that stable population states should be characterized by stable compositions of the female but also male subpopulation, not only by average strategy of females as, for example, in basic game theoretic approaches (see Introduction section in [6] for detailed discussion). In two previous papers [6,7], an alternative synthetic approach to modeling sex ratio evolution was introduced. The new model focuses on the dynamics of sex ratio evolution and tries to combine the phenotypic strategic approach with rigorous genetic analysis. This means that it explicitly considers parental male individuals as the passive carriers of female sex ratio strategies which can be expressed by their daughters if they inherit the strategy from the father. The first formulation [6], is focused on the trajectories of global population parameters such as the primary and the secondary sex ratios and showed that sex ratio evolution dynamics are biphasic. This is caused by fact that equations describing male subpopulation and the secondary sex ratio converge to their stable manifolds faster than female subpopulation. During the first, rapid phase, the population converges to equality between the primary and the secondary sex ratios. Here we are approaching the topic of this paper. During the rapid phase, the male subpopulation is attracted by the unique state, conditional on the current state of the female subpopulation, called the *male subpopulation equilibrium* (thus the dynamics converge to the MSE manifold. See sections 2, 3 and 5.3 in [6] for details). This phenomenon is caused by the mechanism of *fitness exchange*, a contribution of an individual to the fitness of individuals of other types by producing newborns of their type (see section 2 and appendix C in [6] for details). This result shows that the role of passive male carriers as fathers, not only as sons, is also very important. This riddle was partially solved in the second paper [7], which is focused on the selection of genes encoding individual strategies. Selection of these genes depends only on sex ratios in carrier subpopulations rather than directly on trait values encoded by such genes (this self-regulation process can be called *Fisherian Mechanism* see section 2 and 3.1 in [7] for details). However, sex ratios in carrier subpopulations are determined by the action of female carriers of these genes and female partners of male carriers randomly drawn from the population in the process called *Tug of War* (see section 3.2 in [7]). This means that the activity of female carriers attracts the sex ratio among carriers of the particular gene to the value of the encoded by that gene and the activity of random female partners attracts it to the actual value of the primary sex ratio. The *Fisherian Mechanism* and *Tug of war* together constitute the process which was termed *double level selection*. The question arise on the role of the MSE phenomenon in *double level selection* .When the MSE state is reached, selection on strategies appears to work according to the values encoded by the genes (this was shown by lemma 2 in [6]). The new theory produces predictions similar to the classical theory; however, the mechanistic interpretation of the MSE phenomenon remains a mystery and will be analyzed in the present paper. Thus the main goal of this paper is to investigate the properties of the dynamic self-regulation mechanisms discovered in [7] (i.e. Fisherian Mechanism, Tug of War) and their relationships in the process of the double level selection when the trajectory reaches the close neighbourhood of the MSE manifold (sections 2 and 3). This will reveal the importance of the MSE mechanism in the dynamics of sex ratio self-regulation. The next step is investigation of the role of the MSE in the adjustment of fitness exchange mechanism (section 4), which will reveal some intriguing properties of this phenomenon (section 5). Those results will allow comparison of the classical Dusing-Fisher-Shaw-Mohler fitness measure with analogous static fitness function prepared according to the new model.

However, it should be emphasized that the MSE is a specific example of the more general problem. The impact of the passive carriers of unexpressed genes (strategies) on the proliferation of those genes (strategies) may appear in many applications of population genetics and evolutionary game theory. Similar dynamic population equilibria can emerge in problems other than sex ratio evolution, where some fraction of individuals do not exhibit inherited traits, but transfer them to their offspring.

 Next subsections will recall the formal details of the model presented in [6] and [7] and briefly summarize the results obtained there. However, please note that for full understanding of the presented results reading the [6] and [7] would be necessary. From section 2 new results will be presented.

## 1.1 Formal details of the model presented in the previous papers

 Assume that a strategies are expressed by female carriers (as in basic game theoretic approaches and in Dusing's model [1]). There are $u$ individual strategies encoding the sex ratio among newborns of the female, described by $P_i \in [0,1]$. There are $x_i$ female and $y_i$ male carriers of the strategy $P_i$ in the population. Therefore, the population consists of $x = \sum_i x_i$ females and $y = \sum_i y_i$ males. Thus, $f = [f_1,...,f_u]$ is the vector of frequencies of strategies of the female subpopulation, and $m = [m_1,...,m_u]$ is an analogous vector for the male subpopulation (where $f_i = \dfrac{x_i}{x}$ and $m_i = \dfrac{y_i}{y}$ ). $P = \dfrac{y}{y+x}$ is the fraction of males in the population (a secondary sex

ratio), and $\sum_j f_j P_j$ is the mean female subpopulation strategy (the primary sex ratio). Assume

that each female produces $k$ offspring according to haploid inheritance (with probability 0.5 for

gene transfer from the focal parent). However, males are gene carriers too, and transfer those

genes to their offspring. The influence of males can be described by the *fitness exchange effect*

(i.e. daughters of male carriers affect the payoff of female carriers and sons of females affect

payoff of males). This allows for derivation of the sex specific payoff functions, which is outlined

below (see [6] Appendix C for detailed derivation). In [6] it was shown that

$W_{mm} = 0.5\left(\sum_j f_j P_j\right)\dfrac{xk}{y}$ is the number of male offspring, and $W_{mf} = 0.5\left(\sum_j f_j\left(1-P_j\right)\right)\dfrac{xk}{y}$ is the

number of female offspring, of the male individual. These functrions are average values of

binomial distribution, where a trial is the production of a newborn (*xk/y* trials for a single male)

and a success is the production of the individual of sex specified by the function (drawn with probability $\sum_j f_j P_j$ for male and $\sum_j f_j\left(1-P_j\right)$ for female newborn), carrying the strategy gene

from focal parental individual (with probability *0.5*). Analogously, $W_{fm} = 0.5(1-P_i)k$ is the

number of female offspring, and $W_{ff} = 0.5 P_i k$ is the number of male offspring, of the female

individual (*k* trials for a single female). Therefore we have the payoff functions for males

andfemales, respectively, as:

$$W_m(P_i, P, f, m) = W_{mm} + \frac{x_i}{y_i} W_{fm} = k\frac{1-P}{2P}\left(\sum_j f_j P_j + \frac{f_i}{m_i} P_i\right) \quad \text{– payoff function of males,} \quad (1)$$

$$W_f\left(P_i, P, f, m\right) = W_{ff} + \frac{y_i}{x_i} W_{mf} = \frac{k}{2}\left(\left(1-P_i\right) + \frac{m_i}{f_i}\left(1-\sum_j f_j P_j\right)\right) \quad \text{– payoff function of females,} \quad (2)$$

describing per capita number of newborns of the same sex and carried strategy than the focal individual. Now we have all elements needed to formulate the multi-population replicator dynamics [8]. In the first paper [6], this took the following form:

$$\dot{f}_i = f_i\big(W_f(P_i, P, f, m) - \overline{W}_f(P, f, m)\big) \quad \text{for} \quad i = (1, ..., u-1)$$
$$\dot{m}_i = m_i\big(W_m(P_i, P, f, m) - \overline{W}_m(P, f, m)\big) \quad \text{for} \quad i = (1, ..., u-1)$$
$$\dot{P} = P\big(\overline{W}_m(P, f, m) - \overline{W}(P, f, m)\big) \, ,$$

where $\overline{W}_m(P, f, m) = \sum_i m_i W_m(P_i, P, f, m)$, $\overline{W}_f(P, f, m) = \sum_i f_i W_f(P_i, P, f, m)$,

$\overline{W}(P, f, m) = P\overline{W}_m(P, f, m) + (1-P)\overline{W}_f(P, f, m)$ are the respective average payoff functions of the male, female and the whole population (see [6] Appendix D for derivation). This leads to the following system of equations (see [6] Appendix E for derivation):

$$\dot{f}_i = k\left(\frac{f_i}{2}(1-P_i) + \left(\frac{m_i}{2} - f_i\right)\left(1 - \sum_j f_j P_j\right)\right) \quad \text{for} \quad i = (1, ..., u-1) \qquad (3)$$

$$\dot{m}_i = \frac{k}{2}\left(\frac{1-P}{P}\right)\left(f_i P_i - m_i \sum_j f_j P_j\right) \qquad \text{for} \quad i = (1, ..., u-1) \qquad (4)$$

$$\dot{P} = k(1-P)\left(\sum_j f_j P_j - P\right). \qquad (5)$$

It was shown that, for biological reasons, we can limit the analysis of the model to values of the primary and the secondary sex ratios over the interval $(0,1)$. The above formulation is similar to population genetics models and is focused on the composition of the whole population. In [7] the model was modified by a change of coordinates where $G = [G_1, ..., G_{u-1}]$ is the vector of gene frequencies in the whole population (where $G_i = \dfrac{x_i + y_i}{x + y}$), and $M = [M_1, ..., M_u]$

(where $M_i = \dfrac{y_i}{x_i + y_i}$) is the vector of the sex ratios in the subpopulations of carriers of the same strategy genes. This allows taking the perspective of a gene as a strategic agent. The fitness function of a gene encoding strategy $P_i$ (interpreted as the average payoff of the adult carrier of the *i*-th gene) described in the above coordinates and discussed in details in section 2 and appendix B in [7], is therefore:

$$W_g(P_i, G, M) = M_i W_m(P_i, G, M) + (1 - M_i) W_f(P_i, G, M) = \frac{k}{2}\left(\frac{(1-P)}{P} M_i + (1 - M_i)\right). \quad (6)$$

Note that fitness function $W_g$ is independent of the value encoded by individual strategy $P_i$. The fitness of a gene is determined by the sex ratio in a subpopulation of its carriers, $M_i$. This leads to an alternative form of replicator equations (see section 2 and appendix C in [7]):

$$\dot{G}_i = G_i\left(W_g(P_i, P, f, m) - \overline{W}(P, f, m)\right) \qquad \text{dynamics of gene frequencies}$$

$$\dot{M}_i = M_i\left(W_m(P_i, P, f, m) - W_g(P_i, P, f, m)\right) \quad \text{dynamics of sex ratios in carriers subpopulations}$$

which take the form:

$$\dot{G}_i = G_i k\left(\frac{1}{2} - P\right)\left(\frac{M_i}{P} - 1\right) \qquad\qquad\qquad \text{for} \quad i = (1, ..., u - 1) \qquad\qquad (7)$$

$$\dot{M}_i = \frac{k}{2}\left(M_i\left(\frac{1-P}{P}\right)\left(\overline{P}_{pr} - M_i\right) + \left(1 - M_i\right)\left(P_i - M_i\right)\right) \qquad \text{for} \quad i = (1, ..., u). \qquad\qquad (8)$$

An analysis of the replicator equations based on the fitness function $W_g$ (see [7] sections 3.1 and 4.2) shows that changes in a gene pool can be described by the following inequalities:

$G_i$ increases when $\quad P < \frac{1}{2}\quad$ and $\quad P < M_i\quad$ or $\quad P > \frac{1}{2}\quad$ and $\quad P > M_i$,

$G_i$ decreases when $\quad M_i < P < \frac{1}{2}\quad$ or $\quad M_i > P > \frac{1}{2}$, $\qquad\qquad\qquad\qquad\qquad (9)$

$G_i$ is constant when $\quad G_i = 0\quad$ or $\quad M_i = P\quad$ or $\quad P = \frac{1}{2}$ .

This process can be described as the *Fisherian mechanism* (see sections 3.1 and 4.2 in [7] for details). Thus gene pool dynamics depends on the current value of the secondary sex ratio $P$ and the sex ratio in the subpopulation of carriers of a gene $M_i$. Parameter $M_i$ is determined by the action of female carriers of strategy $P_i$ as well as partners of male carriers of this strategy in a so-called *tug of war* mechanism, which means that the activity of female carriers attracts $M_i$ to the value encoded by the gene $P_i$ and the action of partners of male carriers attracts $M_i$ to the value of the females' average strategy $\overline{P}_{pr}$.

## 1.2 The male subpopulation equilibrium phenomenon

The role of the male subpopulation composition was shown by population genetic models of sex ratio evolution [2,3,4,5]. In the MSE state (see sections 2.2 and 5.2 in [6] for details), the condition

$$f_i P_i = m_i \overline{P}_{pr} \tag{10}$$

is satisfied, and the subpopulation of males is in the state $m^{MSE} = \left[ f_1 \dfrac{P_1}{\overline{P}_{pr}}, ..., f_u \dfrac{P_u}{\overline{P}_{pr}} \right]$. The dynamics of the female subpopulation are then described by the rules (according to Lemma 1 in [6]):

$f_i$ increases when $\overline{P}_{pr} < \frac{1}{2}$ and $\overline{P}_{pr} < P_i$ or $\overline{P}_{pr} > \frac{1}{2}$ and $\overline{P}_{pr} > P_i$,

$f_i$ decreases when $P_i < \overline{P}_{pr} < \frac{1}{2}$ or $P_i > \overline{P}_{pr} > \frac{1}{2}$, $\tag{11}$

$f_i$ is stable when: $f_i = 0$ or $f_i = 1$ or $P_i = \overline{P}_{pr}$.

Note that this mechanism is similar to the Fisherian mechanism of evolution of gene frequencies (9). This suggests that the role of the MSE is important in the process of sex ratio self-regulation. The influence of male carriers is more important for the behavior of the system than the value of the secondary sex ratio $P$. The MSE is described by the unclear, nonintuitive condition $f_i P_i = m_i \overline{P}_{pr}$ (see section 4.4 in [7]). This paper will attempt to elucidate the nature of the MSE and specify how it affects the Fisherian and "tug of war" mechanisms. Thus, the next section starts the presentation of the new results.

## 2. The relationship between the MSE and the "tug of war" mechanism

As shown in the previous article [7] (sections 3.2 and 4.3 there), the sex ratio of the subpopulation of carriers of a given gene is self-regulated by the so-called "tug of war" mechanism mentioned here at the end of section 1.1 and this mechanism is described for each strategy by the equation:

$$\dot{M}_i = \frac{k}{2}\left( M_i\left(\frac{1-P}{P}\right)\left(\overline{P}_{pr} - M_i\right) + \left(1 - M_i\right)\left(P_i - M_i\right)\right) \text{ for } \quad i = (1,...,u) .$$  (8)

How does the MSE phenomenon determines equilibria and the dynamics of the "tug of war" process? The solution to this problem is described below by the following lemmas. Consider first how the MSE affects the sex ratio in the carrier subpopulation $M_i$ :

**Lemma 1**

When the MSE condition (10) for strategy $P_i$ is satisfied, then according to the current value of $\overline{P}_{pr}$, we have:

$$\overline{\overline{M}}_i = \frac{P_i}{\Gamma \overline{P}_{pr} + P_i} \qquad \left( \overline{\overline{M}}_i = \frac{P_i}{1 - \overline{P}_{pr} + P_i} \quad \text{when } \overline{P}_{pr} = P \right)$$  (12)

and if the MSE condition is satisfied for all $i$, then the condition $\overline{P}_{pr} = P$ can be maintained despite changes in gene frequencies $G_i$.

For proof see Appendix 1.

Therefore, the MSE state is equivalent to the unique sex ratio in the carrier subpopulation $\overline{\overline{M}}_i$, and it is a necessary condition to make equality of the primary and the secondary sex ratios stable despite changes in gene frequencies. The question arises about the role of $\overline{\overline{M}}_i$ in the "tug of war" process.

**Lemma 2**

a) $\overline{\overline{M}}_i$ is not a stationary point of the dynamics (8), but for the current value of $\overline{P}_{pr}$, the dynamics is attracted by the point

$$\hat{M}_i = \frac{2P_i\overline{P}_{pr}}{P_i + \overline{P}_{pr}} \, .$$ (13)

b) If $\overline{P}_{pr} = 0.5$ and $P = \overline{P}_{pr}$ then $M_i = \hat{M}_i$ for all $i$, which means that all subpopulations of carriers are in stable states.

For a proof see Appendix 2.

The value of $\overline{\overline{M}}_i$ determined by the MSE of the individual strategy $P_i$, is very important. This parameter is an argument of the fitness function of a gene $W_g$ (6). Different strategies converge to different values of $\overline{\overline{M}}_i$, and through these values, they can be distinguished by natural selection operating on a second, higher level of double level selection. In addition, the male subpopulation equilibrium is not equivalent to certain stable sex ratios in the carrier subpopulation, but it affects the evolution of this parameter by determining the value of the attractor $\hat{M}_i$. In a sense, $\overline{\overline{M}}_i$ is a "wolf" that chases a "rabbit" $\hat{M}_i$ during the run on the MSE manifold. "Wolves" of all subpopulations catch their rabbits at the global equilibrium point.

## 3. The influence of the MSE on gene frequencies and the primary sex ratio

Another question addresses the influence of the male subpopulation equilibrium on gene frequencies. The lemma described below shows that this influence plays a crucial role in the phenomenon of sex ratio self-regulation. When the MSE condition is satisfied then the dynamics (9) is driven by encoded values of individual strategies. The lemma below describes the relationship between gene frequency $G_i$ and the frequency of its carriers among females $f_i$:

**Lemma 3**

If the MSE condition (10) is satisfied, $P = \overline{P}_{pr}$ and population is in the polymorphic state, then:

$$G_i = f_i(P_i + 1 - \overline{P}_{pr}).\tag{14}$$

Parameter $G_i$ changes according to the following rules. For $G_i > 0$:

$G_i$ increases when $\overline{P}_{pr} < \dfrac{1}{2}$ and $\overline{P}_{pr} < P_i$ or $\overline{P}_{pr} > \dfrac{1}{2}$ and $\overline{P}_{pr} > P_i$

$G_i$ decreases when $P_i < \overline{P}_{pr} < \dfrac{1}{2}$ or $P_i > \overline{P}_{pr} > \dfrac{1}{2}$ \hfill (15)

and

$G_i$ is stable when $G_i = 0$ or $P_i = \overline{P}_{pr}$ or $\overline{P}_{pr} = \dfrac{1}{2}$.

For a proof see Appendix 3.

This result shows that the male subpopulation equilibrium determines the relationship between changes in a gene pool and frequencies of strategies in the female subpopulation. Therefore, changes in gene frequencies are synchronized with changes in the mean female subpopulation strategy (the primary sex ratio $\overline{P}_{pr}$). The dynamics of $G_i$ are similar to the dynamics of $f_i$ ((11), see also Lemma 1 from [6]) and have the same stationary points. This is explained by the relationship (14).

## 4. The effect of the MSE on the sex-specific payoff functions (1) and (2) and the "fitness exchange" effect

Let us examine the form of payoff functions if a population is in the MSE state.

Recall that male and female payoff functions (1) and (2) describe numbers of newborns of the same sex and carrying the same strategy than the focal individual. Lemma 4 describes the quantitative properties of the "fitness exchange" mechanism in the MSE state and its impact on sex-specific payoffs.

**Lemma 4**

When the MSE condition is satisfied and $\overline{P}_{pr} = P$ then:

(i) $\dfrac{x_i}{y_i} W_{fm} = \dfrac{k}{2} \Gamma \overline{P}_{pr}$ and a male's payoff (1) takes the form $W_m = k\Gamma \overline{P}_{pr}$

(ii) $\dfrac{y_i}{x_i} W_{mf} = \dfrac{k}{2} \Gamma P_i$ and a female's payoff (2) has the form $W_f = \dfrac{k}{2}\big((1 - P_i) + P_i\Gamma\big)$

(iii) MSE condition (10) is equivalent to

$$x_i\, P_i k = y_i \Gamma \overline{P}_{pr} k \,,$$

which means that all female carriers of gene $P_i$ produce the same number of sons as partners of all male carriers of that gene.

For a proof, see Appendix 4

However, above lemma shows not only the forms of sex-specific payoffs on the MSE manifold. It reveals some interesting relationships. The reproductive values of the female and male carries are $\dfrac{k}{2}$ and $\dfrac{k}{2}\Gamma$. Point (i) shows that the female carriers of a gene produce $\dfrac{k}{2}\Gamma \overline{P}_{pr}$ new male carriers per single adult male carrier of this gene (the same as his own partners (mates)). In effect, the payoff of every male in the population, irrespective of the carried strategy, is equal to $k\Gamma \overline{P}_{pr}$.

Thus the reproductive value of the "genetic sons" (carrying the same strategy than a focal parent)

of the focal male equals the reproductive value of the contribution of female carriers to his payoff.

Similarly, male carriers of a gene produce $\frac{k}{2}\Gamma P_i$ new female carriers per single adult female carrier of this gene, leading to (ii). Note that the reproductive value of "genetic nieces" of the focal female, produced by male carriers is $\frac{k^2}{4}\Gamma P_i$ and equals the reproductive value of the "genetic sons" of that female (since she produces $\frac{k}{2}P_i$ male carriers). Thus:

**Corollary 1**

In the MSE state, the reproductive value of the contribution of the opposite sex gene carriers to the payoff of the focal individual equals the reproductive value of the newborn male carriers produced by that individual.

Point (iii) provides a more intuitive interpretation of point (i), showing that in the MSE state, evolution of the carriers' subpopulations sex ratio $M_i$ is driven by the fact that both groups of all female carriers and partners of all male carriers produces the same number of sons but different numbers of daughters. Every single male in the population produces the same number of daughters, despite carried strategies. However, different proportions $\frac{y_i}{x_i}$ for different strategies cause differences in the influence of males on female payoffs. This pattern appears to be an effect of the inheritance system (in this case haploid).

## 5. The relationship between global and carrier subpopulation parameters.

Lemma 4 shows that the payoff of each male individual is equal, irrespective of his carried strategy. This leads to another surprising property of the MSE and the "fitness exchange" mechanism that affects not only individuals carrying the same strategy, but the population as a whole. Similarly to point (iii) of Lemma 4, since $f_i = \dfrac{x_i}{x}$ and $m_i = \dfrac{y_i}{y}$, the MSE condition (10) is equivalent to:

$$\frac{x}{y} \sum_j f_j P_j = \frac{x_i}{y_i} P_i \qquad \left( \Gamma \overline{P}_{pr} = \Gamma_i P_i \right).$$   (16)

This means that the product of the number of females per single male $\dfrac{x}{y}$ and the primary sex ratio in the entire population $\sum_j f_j P_j$ is equal to the analogous coefficients for monomorphic subpopulations of carriers of any gene $P_i$. This property can be generalized.

Consider the subpopulation described as $\Lambda$, consisting of all carriers of arbitrarily chosen strategies described by the vector of indices $d^\Lambda = [d_1^\Lambda, ..., d_{u^\Lambda}^\Lambda]$ (for example $[1,3,4]$ means, that carriers of strategies $P_1, P_3, P_4$ are chosen). Assume that $\Gamma^\Lambda$ is the number of females per single male individual, and $\overline{P}_{pr}^\Lambda$ is the mean strategy of females from subpopulation $\Lambda$ (i.e., the subpopulation primary sex ratio). Then point (iii) from Lemma 4 can be generalized to the interesting property described by the following lemma.

**Lemma 5**

When the MSE condition is satisfied, then for subpopulation $\Lambda$, the following condition is satisfied:

$$\Gamma \overline{P}_{pr} = \Gamma^\Lambda \overline{P}_{pr}^\Lambda,$$   (17)

In particular, $\Gamma \overline{P}_{pr} = \Gamma^i P_i$ for a monomorphic subpopulation of carriers of strategy $P_i$ for which $\Gamma^i = \dfrac{x_i}{y_i}$ .

For a proof see Appendix 5.

The result obtained above has an interesting application. Condition $\Gamma \overline{P}_{pr} = \Gamma^\Lambda \overline{P}_{pr}^\Lambda$ allows the estimation of the primary sex ratio of the entire population by parameters of an arbitrarily chosen subpopulation. There is no need to know all strategies in the population. We need only one genotype (or a few) encoding an individual strategy. By analyzing a subpopulation of carriers of these genes, we can estimate the primary sex ratio of the whole population as

$$\overline{P}_{pr} = \frac{\Gamma^\Lambda \overline{P}_{pr}^\Lambda}{\Gamma} .  \tag{18}$$

## 6. Comparison of the new theory with classical Dusing-Fisher-Shaw-Mohler approach

This section examines the results of the classical theory according to the new approach. Numerical solutions for the new model suggest that both approaches produce compatible predictions, if the male subpopulation equilibrium condition (MSE) is satisfied. However, there is no formal proof supporting these observations. In addition, the interpretation of the classical fitness measure as a Malthusian growth rate is problematic from the perspective of the new model. The classical static fitness measure (described as DFSM fuction in [6]) relies on the number of grand offspring of a female, which expresses her individual strategy:

$$W(P_i, P) = \tfrac{k^2}{4}(1-P)\left(\frac{1-P_i}{1-P} + \frac{P_i}{P}\right) = \tfrac{k^2}{4}\left(1 - P_i + \frac{1-P}{P}P_i\right) =$$

$$= \frac{k^2}{4}\left(\left(\frac{1-P}{P}-1\right)P_i+1\right) = \frac{k^2}{4}\left(1-P_i+\frac{1-P}{P}P_i\right) = \frac{k^2}{4}\left((1-P_i)+P_i\Gamma\right). \qquad (19)$$

Note that according to point (ii) of Lemma 4 and Corollary 1 we have:

$$W(P_i, P) = W_f \frac{k}{2}.$$

Thus the reproductive value of the male and female offspring of the focal female is equal to the reproductive value of the female carriers produced by the focal female and her "genetic brothers". Both fitness measures $W(P_{ind}, P)$ and $W_f \frac{k}{2}$ are biased and disregard the sons of the male carriers.

In addition $W(P_{ind}, P)$ also disregards the daughters of male carriers while $W_f \frac{k}{2}$ disregards the sons of the focal female. However, according to the point (ii) of the Lemma 4 and the Corollary 1, the impact of those factors on the female payoff is equivalent, which leads to the equivalence of $W(P_{ind}, P)$ and $W_f \frac{k}{2}$ as the fitness measures.

The gene fitness function is $W_g = \frac{k}{2}\left(\Gamma M_i + (1-M_i)\right)$, where $M_i$ is the sex ratio in the carrier subpopulation of the $i$-th strategy and $\Gamma = \frac{1-P}{P}$ describes the mean number of female partners of a male individual. Below is the analysis of the strategic equivalence of $W$ and $W_g$ when the MSE condition is satisfied.

**Lemma 6: On strategic compatibility of fitness measures**

If the MSE condition is satisfied for all strategies, then the gene fitness function for any arbitrarily chosen strategies $P_1$ and $P_2$ has the form:

$$W_g(P_i, G, M) = \frac{k}{2}\left(\frac{\overline{P}_{pr} + P_i}{\overline{P}_{pr} + \dfrac{P_i}{\Gamma}}\right) \tag{20}$$

and the following condition is satisfied:

$W(P_1, P) \le W(P_2, P)$ if and only if $W_g(P_1, G, M) \le W_g(P_2, G, M)$,

which means that both the new and classical fitness measures are compatible with regard to the relation of payoff dominance of strategies.

For a proof, see Appendix 6.

Note that in Lemma 6, only the MSE condition needs to be satisfied so that both fitness functions can produce compatible strategic predictions. The secondary sex ratio $P$ may be far from equilibrium (which means $\overline{P}_{pr} \neq P$). Therefore both functions are compatible on the level of static game theoretic analyses and the MSE is a sufficient condition for this. The question arises about the equivalence of both approaches on the grounds of population dynamics.

If we assume that $P = \overline{P}_{pr}$, then the fitness function of a gene will have the form:

$$W_g(P_i, G, M) = \frac{k}{2}\left(\frac{\overline{P}_{pr} + P_i}{\overline{P}_{pr} + \dfrac{P_i}{\Gamma}}\right) = \frac{k}{2}\left(\frac{\overline{P}_{pr} + P_i}{\overline{P}_{pr} + \dfrac{\overline{P}_{pr} P_i}{1 - \overline{P}_{pr}}}\right) = \frac{k}{2}\left(\frac{1 + \dfrac{P_i}{\overline{P}_{pr}}}{1 + \dfrac{P_i}{1 - \overline{P}_{pr}}}\right).$$

Let us assume for simplicity that the mean brood size of a female is $k = 2$, which will remove this parameter from both formulae. In effect, two functions of parameters $P_i$ and $\overline{P}_{pr}$ are obtained:

$$W_g = \left(1 + \frac{P_i}{\overline{P}_{pr}}\right) \bigg/ \left(1 + \frac{P_i}{1 - \overline{P}_{pr}}\right) \quad \text{and} \quad W = \frac{1 - 2\overline{P}_{pr}}{\overline{P}_{pr}} P_{ind} + 1 \ .$$

Now the fitness measures can be compared. Fig. 1 shows the plots of both functions and the reminder $W - W_g$. It is clear that these functions are similar but not identical. Fig. 2 shows cross-sections of surfaces from Fig. 1 near the equilibrium. The classical fitness measure is a linear function of parameter $\overline{P}_{pr}$, whereas the new fitness measure is nonlinear. Despite disregarding the effect of male carriers, Lemma 6 shows that the classical fitness measure related to the number of grand offspring is a good approximation of reproductive success according to strategic analyses if the MSE condition is satisfied. However, Figures 1 and 2 show that this is a biased approximation, which may have serious implications when modeling detailed population dynamics.

Suppose that a mutant female produces all carriers of a mutant gene $P_2$ in the next generation according to the expression of her individual strategy. Thus, in the generation of her descendants, there is equality $M_2 = P_2$, and the reproductive success of a mutant gene in the first generation of descendants is proportional to $\left( \Gamma P_2 + \left( 1 - P_2 \right) \right)$. It is clear that the predictions of this function are exactly the same as in a DFSM model. This also supports the compatibility of linearization of a limit case of the population genetics model [5] with the DFSM approach. A problem arises in further generations. For each female carrying mutant gene $P_2$, there is some number of male carriers of unexpressed gene $P_2$, acting as fathers. The DFSM fitness measure, which is related to the number of a female's grand offspring, disregards the role of male carriers on gene proliferation. Thus the classical fitness measure can be seriously biased in cases where an entire subpopulation carries a mutant gene, not only a single female (see Fig. 3). This bias may be strong for strategies close to 1 that produce mostly males.

## 7. Discussion

## 7.1 The role of the MSE in sex ratio self-regulation

Let us summarize the obtained results. The MSE plays a crucial role in sex ratio self-regulation by affecting the following mechanisms:

**- the "Fisherian" mechanism:**

According to Lemma 1, the state of the male subpopulation equilibrium (in which all individuals from the population are produced by females from this population) is equivalent to the unique value of the sex ratio in the carrier subpopulation, equal to $\overline{\overline{M}}_i = \dfrac{P_i}{\Gamma \overline{P}_{pr} + P_i}$. This is the connection between the value of the coded strategy $P_i$ with the payoff of a gene that depends only on the sex ratio in the carrier subpopulation $M_i$, and the secondary sex ratio $P$. Through parameter $\overline{\overline{M}}_i$, the individual strategy $P_i$ may affect the "Fisherian" mechanism.

**- the "Tug of War" mechanism:**

The value $\overline{\overline{M}}_i$ is not a stationary point in the "tug of war" dynamics, but for the current value of $\overline{P}_{pr}$, the dynamics is attracted by the point $\hat{M}_i = \dfrac{2P_i \overline{P}_{pr}}{P_i + \overline{P}_{pr}}$, which is also determined by the state of the male subpopulation equilibrium.

**-the "fitness exchange" mechanism:**

The MSE is the state where all female carriers of gene $P_i$ produce the same number of sons as partners of all male carriers of that gene (Lemma 4). The reproductive value of the contribution of the opposite sex gene carriers to the payoff of the focal individual equals the reproductive value of her\his "genetic" sons (Corollary 1).

**- the relationship between gene frequencies and the primary sex ratio:**

Another important relationship that is induced by the male subpopulation equilibrium is the equality $G_i = f_i(P_i + 1 - \overline{P}_{pr})$. This condition, presented in Section 3 (see Lemma 3), links changes in gene frequencies affected by the "Fisherian mechanism" with changes in the primary sex ratio (the female subpopulation mean strategy). This is the missing element closing the feedback loop of population self-regulation. In addition for every subpopulation consisting of all carriers of some genes the product of a subpopulation's primary sex ratio and the number of females per single male equals the analogous parameter for the whole population (Lemma 5).

**-relationship between classical and the new theory**

In the MSE state the DFSM fitness measure is equivalent to the sex specific female payoff $W_f$ and both of those fitness measures are biased approximations of the gene fitness functions $W_g$. The bias increases with the fraction of males among the offspring (see Fig. 1 and 2).

## 7.2 The relationship between subpopulation parameters and global population parameters induced by the MSE

The first important property is that the MSE state is necessary in all carrier subpopulations to maintain equality between the primary and the secondary sex ratios during changes in gene frequencies. However, there is another interesting property produced by the MSE phenomenon. When the MSE condition is satisfied for all individual strategies, then the condition $\Gamma \overline{P}_{pr} = \Gamma^i P_i$ is satisfied. In other words, the product of the mean female subpopulation's strategy (the primary sex ratio) and the number of females per single male in the whole population is equal to the product of the strategy value $P_i$ and the number of females per single male in the subpopulation of the carriers of $P_i$ (denoted by $\Gamma^i$). This result is generalized in Lemma 5, which shows that this equality holds for the analogous parameter calculated for any subpopulation consisting of all carriers of some arbitrarily chosen subset of individual strategies. This property reveals that to approximate the primary sex ratio of the entire population, it is sufficient to know the primary and secondary sex ratios of an arbitrarily chosen subpopulation and the secondary sex ratio of the whole population. Empirical measurement of the sex ratio among zygotes is complicated, especially in a population highly diversified by individual strategies. In contrast, the secondary sex ratios (in effect, parameters $\Gamma^\Lambda$ and $\Gamma$) are easy to measure. Therefore, to calculate the primary sex ratio, we only need to identify a single gene (or several genes) that encodes an individual strategy and measure the primary sex ratio in the subpopulation of carriers of this gene (or genes) $\overline{P}_{pr}^{\Lambda}$. At this point, it becomes possible to estimate the primary sex ratio of the entire population by $\overline{P}_{pr} = \dfrac{\Gamma^\Lambda \overline{P}_{pr}^{\Lambda}}{\Gamma}$. The advantage of this approach is that it does not require information about all individual strategies in the population. However, the usefulness of this method is questionable because there is no information about the robustness of an MSE against differences in the costs of producing sons and daughters. The MSE condition can be derived because the mean number of offspring $k$ is equal for every female, which implicitly assumes equal costs of producing a single offspring, irrespective of sex. If we extend the model to dependence on costs and resources, then parameter $k$ will depend on the individual strategy $P_i$. In effect, the MSE equilibrium will have a different form or may not exist, a possibility that requires further studies.

## 7.3 The Dusing-Fisher-Shaw-Mohler model from the perspective of the new approach

It was shown that in the MSE state, the same function describes the reproductive value of the offspring of the focal female and the reproductive value of the female newborns produced per single female carrier. Thus, DFSM fitness measure is equivalent to the reproductive value of the sex specific female payoff. Both approaches are biased approximations of the new approach related to the gene perspective. However, they are compatible with the new approach at the level of the strategic analyses and produce the same game theoretic predictions. A disadvantage of the classical sex ratio game is the fact that it disregards passive male gene carriers as fathers mating with other females (note that equivalent fitness measure disregards the impact of the male offspring). They are considered only as the sons of "mom's" ([9]). The strategic agent is a female individual. This produces a few problems. The first is a bias in estimation of the reproductive success which is problematic on the grounds of population dynamics. This is an example of the disadvantage of purely strategic models mentioned by [10] (more on this topic in [11,12]). In effect the basic interpretation of the payoff as the Malthusian growth rate is problematic in the case of the classical sex ratio game (dynamic models based on the Dussing's approach can be found for example in [13]). This is an important problem, because the sex ratio game is a basic example of a nonlinear payoff function in every textbook on evolutionary game theory. A second problem is that the evolutionarily stable equilibrium (sex ratio of 0.5) is described as a state of the female subpopulation. The male subpopulation is not explicitly considered. Classical population genetics results show that the state of the male subpopulation is important and the sex ratio of 0.5 can be unstable for perturbations of that state [2,3,4,5]. The new model also supports these predictions. In the new approach the strategic agent is the gene encoding the sex ratio strategy. This perspective is free of the above disadvantages and leads to new interesting predictions such as the "tug of war" mechanism and double level selection. It is also consistent with classical population genetics results.

## 7.4 The mechanistic nature of the MSE

Lemma 4 (see section 4) shows that in the MSE state, a group of male gene carriers produce the same number of genetic sons as the group of partners of these male carriers. In addition, the payoff of every male in the population is equal to $k\Gamma\overline{P}_{pr}$, and the payoff of a female with strategy $P_i$ is equal to $\frac{1}{2}k\left((1-P_i)+P_i\Gamma\right)$. Therefore, the influence of female carriers on the payoff of a male carrier (described by component $\frac{x_i}{y_i}W_{fm}(P_i)$ of payoff function (1)) is in equilibrium (the same for all and equal to $\frac{k}{2}\Gamma\overline{P}_{pr}$), whereas the influence of male carriers on the payoff of a female is not in equilibrium (it depends on the individual strategy and is equal to $\frac{k}{2}\Gamma P_i$). The mechanistic explanation of this process in the male subpopulation is simple. The reproductive activity of male carriers is selectively neutral. Partners of every male produce the same number of offspring ($\frac{k}{2}\Gamma\overline{P}_{pr}$), so the differences in payoff between males carrying different genes correspond to the effects of differences in their "sisters'" activities (female carriers of the same gene). The influence of female carriers is determined by the factor $\frac{x_i}{y_i}$ (the number of female carriers per single male carrier). If this factor has a high value (more females per single male), then male carriers of this gene will experience a greater growth rate. However, when the number of male carriers $y_i$ increases, then the factor $\frac{x_i}{y_i}$ will decrease (see Fig. 4). This will cause a decrease in the influence of female carriers on the payoff of the average male carrier. This process will lead to an equilibrium state in which the influence of female carriers for all male carriers is the same. The question arises: How does this mechanism affect the female subpopulation? Since the influence of a single female on the fitness of a male $W_{fm}(P_i)$ depends on the carried strategy, the equilibrium mechanism described above implies different values of factor $\dfrac{x_i}{y_i}$ that describe the per capita number of females per single male. The influence of a single male on the fitness of a female $W_{mf}$ is independent of the carried strategy. Differences in the contributions of male carriers to female fitness are determined by different values of the coefficient $\dfrac{y_i}{x_i}$ (describing the number of male carriers per single female) for different strategies. Female carriers of different strategies exert continual pressure on the equilibrium state of a male subpopulation by introducing different numbers of new male carriers. In effect, through this subtle population mechanism, female individuals indirectly influence the feedback of male carriers on their own fitness and thus control the contribution of male carriers to their own fitness (Lemma 4 (ii)). In addition, this outcome leads to different sex ratio values in carrier subpopulations when the entire population is in the MSE.

The results presented above attempt to interpret the MSE phenomenon. They suggest that the MSE constitutes a compensatory equilibrium between the production of female carriers of a gene by male carriers and the production of male carriers by female individuals (*a fitness exchange effect*). As was mentioned in the introduction, the MSE is an example of a more general class of problems. The MSE is related to population phenomena, such as stable demographic structure or Hardy-Weinberg equilibrium. In the new model, there is no demographic structure. However, existing models that include demographic structure [14] report different behaviors before and after demographic equilibrium is reached. The results from this paper also suggest that the male subpopulation equilibrium is the state to which the population rapidly converges when all individuals in the population are produced only by females from this population (no migration or other external factors). The MSE properties presented in this paper represent effects of haploid inheritance and the assumption that an individual strategy is encoded by a single gene. However, other forms of self-regulation structures are possible for different inheritance systems (diploid or haplodiploid) and different genetic structures (multilocus or polygenic). Here strategic analysis of phenotypic adaptation meets population genetics. Strictly genetic mechanisms may play important roles in the process of phenotypic selection by determining the values of crucial parameters (e.g., the sex ratio in a carrier subpopulation) that are responsible for the selection of individual strategies. Similar dynamic equilibria should be observed in every model where there are passive carriers of unexpressed strategy genes.

This work was supported by grant N N304 2764 33 from the Polish Ministry of Science and Higher Education.

References

[1] A.W.F. Edwards,. Carl Dusing (1884) on the Regulation of the Sex-Ratio, Theor. Pop. Biol 58 (2000) 255-257.

[2] I. Eshel, M. Feldman, On Evolutionary Genetic Stability of the Sex Ratio, Theor. Pop. Biol 21 (1982) 430-439.

[3] I. Eshel, M. Feldman, On the Evolution of Sex Determination and the Sex Ratio in Haplodiploid Populations, Theor. Pop. Biol 21 (1982) 440-450.

[4] S. Karlin, S. Lessard, Theoretical Studies on Sex Ratio evolution, Princeton University Press, 1986.

[5] J. Seger, J.W. Stubblefield,. Models of Sex Ratio Evolution. In: Hardy, I.C.W. (ed.). Sex

Ratios, Concepts and Research Methods, Cambridge University Press, 2002.

[6] K.Argasinski "The dynamics of sex ratio evolution : The dynamics of self-regulation of a population state" J. Theor. Biol. 309 (2012) 134–146

[7] K.Argasinski "The dynamics of sex ratio evolution : From the gene perspective to multilevel selection" PLOS One  10.1371/journal.pone.0060405

[8] K.Argasinski Dynamic multipopulation and density dependent evolutionary games related to replicator dynamics. A metasimplex concept.  Math. Biosci. 202, 88-114 2006

[9] Charnov, E.L., 1982. The Theory of Sex Allocation. Princeton University Press.

[10] G.F. Oster, S.M. Rocklin, Lectures on Mathematics in the Life Sciences, Vol XI, ed. S.A. Levin, AMS, 1979, pp. 21-88.

[11] Bomze, IM, Shuster P, Sigmund K, (1983) The role of Mendelian genetics in strategic models on animal behavior. J. Theor. Biol. 101, 19–38.

[12] Weissing FJ (1996) Genetic versus phenotypic models of selection: can genetics be neglected in a long-term perspective? J. Math. Biol. *34*(5), 533-555.

[13] MacArthur, R. H., T. H. Waterman, and H. Morowitz. Theoretical and Mathematical Biology. Blaisdell, New York (1965): 388.

[14] P.D. Taylor, A general mathematical model for sex allocation, J. Theor. Biol. 112 (1985), 799-818.

## Appendix 1

## Proof of Lemma 1

Assume that the male subpopulation equilibrium condition $m_i \overline{P}_{pr} = f_i P_i$ is satisfied. Recall that

$$M_i = \frac{P m_i}{G_i} \quad \left( m_i = \frac{M_i G_i}{P} \right) \text{ and } 1 - M_i = \frac{(1-P) f_i}{G_i} \quad \left( f_i = \frac{(1-M_i) G_i}{1-P} \right). \tag{21}$$

Then the MSE condition can be written in the form:

$$\frac{M_i G_i}{P} \overline{P}_{pr} = \frac{(1 - M_i) G_i}{1 - P} P_i$$

then:

$$M_i = \frac{P(1 - M_i)}{1 - P} \frac{P_i}{\overline{P}_{pr}} = \frac{(1 - M_i)}{\Gamma} \frac{P_i}{\overline{P}_{pr}} = \frac{P_i}{\Gamma \overline{P}_{pr}} - M_i \frac{P_i}{\Gamma \overline{P}_{pr}},$$

thus

$$M_i \left( 1 + \frac{P_i}{\Gamma \overline{P}_{pr}} \right) = \frac{P_i}{\Gamma \overline{P}_{pr}},$$

which gives

$$M_i \left( \frac{\Gamma \overline{P}_{pr} + P_i}{\Gamma \overline{P}_{pr}} \right) = \frac{P_i}{\Gamma \overline{P}_{pr}}.$$

The male subpopulation equilibrium condition is therefore equivalent to the following sex ratio value in the carrier subpopulation:

$$\overline{\overline{M}}_i = \frac{P_i}{\Gamma \overline{P}_{pr} + P_i} \qquad \left( \overline{\overline{M}}_i = \frac{P_i}{1 - \overline{P}_{pr} + P_i} \quad \text{when } \overline{P}_{pr} = P \right).$$

The last question addresses when equality $\overline{P}_{pr} = P$ can be maintained despite changes in gene frequencies $G_i$. Recall that $\overline{P}_{pr} = \frac{1}{1 - P} \sum_j (1 - M_j) G_j P_j$ and $P = \sum_j G_j M_j$ (see section 2 in [7]).

Equality of the primary and the secondary sex ratios therefore has the form:

$$\frac{1}{1 - P} \sum_j (1 - M_j) G_j P_j = \sum_j G_j M_j.$$

The above equality will be independent of the value of $G_j$ if for all $j$, equality between the $j$-th coefficients of the sum is satisfied. Then $G_j$ reduces. In effect, we obtain:

$$(1 - M_j)\frac{P_j}{1 - P} = M_j.$$

From this formula we obtain:

$$M_j = \frac{P_j}{1 - P + P_j}.$$

After substituting $\overline{P}_{pr} = P$, we obtain condition on $M_j$, which is equivalent to the sex ratio in the carrier subpopulation equivalent to the MSE state. This means that MSE state is necessary condition to maintain equality between the primary and the secondary sex ratios during changes in gene frequencies.

## Appendix 2

## Proof of Lemma 2

Because $f_i = \frac{x_i}{x}$ and $m_i = \frac{y_i}{y}$, according to the MSE condition (10), we have

$$\frac{\Gamma}{\Gamma^i} = \frac{P_i}{\overline{P}_{pr}}, \tag{22}$$

where $\Gamma^i = \frac{1 - M_i}{M_i} = \frac{x_i}{y_i}$ and $\Gamma = \frac{1 - P}{P} = \frac{x}{y}$. The equation for the sex ratio in the carrier subpopulation (8) has the form:

$$\dot{M}_i = \frac{k}{2}\left( M_i(\overline{P}_{pr} - M_i)\Gamma + (1 - M_i)(P_i - M_i) \right).$$

For every value of $M_i$ from the interior of the unit interval, the right hand side of this equation can be presented in the form:

$$\frac{k}{2}(1 - M_i)\left( \frac{M_i}{(1 - M_i)}(\overline{P}_{pr} - M_i)\Gamma + (P_i - M_i) \right) = \frac{k}{2}(1 - M_i)\left( \frac{\Gamma}{\Gamma^i}(\overline{P}_{pr} - M_i) + (P_i - M_i) \right). \tag{23}$$

Thus by substitution of (22), we can transform (23) to:

$$\frac{k}{2}\left(1-M_i\right)\left(\frac{P_i}{\overline{P}_{pr}}\left(\overline{P}_{pr}-M_i\right)+\left(P_i-M_i\right)\right).$$

Let us find the zero point from the interior of the unit interval of the right side of the equation transformed to the above formula:

$$\frac{P_i}{\overline{P}_{pr}}\left(\overline{P}_{pr}-M_i\right)+P_i-M_i=0.$$

Then

$$P_i-\frac{P_i}{\overline{P}_{pr}}M_i+P_i-M_i=0,$$

thus

$$2P_i-M_i\left(\frac{P_i}{\overline{P}_{pr}}+1\right)=0,$$

thus

$$2P_i=M_i\left(\frac{P_i+\overline{P}_{pr}}{\overline{P}_{pr}}\right).$$

In effect, we obtain the zero point:

$$\hat{M}_i=\frac{2P_i\overline{P}_{pr}}{P_i+\overline{P}_{pr}}.$$

Thus point a) is proven.

Substitution of $\overline{P}_{pr}=0.5$ and $P=\overline{P}_{pr}$ implies:

$$\overline{\overline{M}}_i=\hat{M}_i \quad \text{for all} \ \ i,$$

which constitutes the end of the proof. $\boxed{\cdot}$

# Appendix 3

# Proof of Lemma 3

$G_i$ can be presented in the form (see [7] section 2):

$$G_i = Pm_i + (1-P)f_i.$$

By substituting $m_i = f_i \dfrac{P_i}{\overline{P}_{pr}}$ (according to MSE condition (10)) and $P = \overline{P}_{pr}$, we obtain:

$$G_i = Pf_i \frac{P_i}{\overline{P}_{pr}} + (1-P)f_i = f_i\left(\frac{P}{\overline{P}_{pr}}P_i + (1-P)\right) = f_i\left(P_i + 1 - \overline{P}_{pr}\right).$$

This is the relationship between $G_i$ and $f_i$. Let us focus on the dynamics of parameter $G_i$ (7):

$$\dot{G}_i = G_i k\left(\frac{1}{2} - P\right)\left(\frac{M_i}{P} - 1\right).$$

The product $\left(\dfrac{1}{2} - P\right)\left(\dfrac{M_i}{P} - 1\right)$ is responsible for the sign of the right side of this equation. We have assumed that the primary and the secondary sex ratios and that the MSE condition (10) is satisfied. Thus, by substituting $P = \overline{P}_{pr}$ and condition (12) $M_i = \overline{\overline{M}}_i = \dfrac{P_i}{1 - \overline{P}_{pr} + P_i}$ to (7), we obtain:

$$\left(\frac{1}{2} - \overline{P}_{pr}\right)\left(\frac{P_i}{\overline{P}_{pr}\left(1 - \overline{P}_{pr} + P_i\right)} - 1\right) = \left(\frac{1}{2} - \overline{P}_{pr}\right)\frac{P_i - (1 - \overline{P}_{pr})\overline{P}_{pr} - \overline{P}_{pr}P_i}{\overline{P}_{pr}\left(1 - \overline{P}_{pr} + P_i\right)} =$$
$$\left(\frac{1}{2} - \overline{P}_{pr}\right)\frac{P_i(1 - \overline{P}_{pr}) - (1 - \overline{P}_{pr})\overline{P}_{pr}}{\overline{P}_{pr}\left(1 - \overline{P}_{pr} + P_i\right)} = \left(\frac{1}{2} - \overline{P}_{pr}\right)\frac{(1 - \overline{P}_{pr})\left(P_i - \overline{P}_{pr}\right)}{\overline{P}_{pr}\left(1 - \overline{P}_{pr} + P_i\right)}.$$

Because $\Gamma = \dfrac{(1-P)}{P} = \dfrac{(1 - \overline{P}_{pr})}{\overline{P}_{pr}}$, The right hand side of equation (7) equals to:

$$\left(\frac{1}{2} - \overline{P}_{pr}\right)\frac{\Gamma\left(P_i - \overline{P}_{pr}\right)}{(1 - P + P_i)}.$$

Because $\Gamma$ and $(1 - P + P_i)$ are always nonnegative, the sign of the formula above describing the right side of equation $\dot{G}_i$ is determined by the product:

$$\left(\frac{1}{2} - \overline{P}_{pr}\right)\left(P_i - \overline{P}_{pr}\right).$$

Therefore,

$G_i$ increases when $\overline{P}_{pr} < \frac{1}{2}$ and $\overline{P}_{pr} < P_i$ or $\overline{P}_{pr} > \frac{1}{2}$ and $\overline{P}_{pr} > P_i$

$G_i$ decreases when $P_i < \overline{P}_{pr} < \frac{1}{2}$ or $P_i > \overline{P}_{pr} > \frac{1}{2}$

$G_i$ is stable when $G_i = 0$ or $P_i = \overline{P}_{pr}$ or $\overline{P}_{pr} = \frac{1}{2}$,

which ends this proof. $\boxed{\cdot}$

## Appendix 4

## Proof of Lemma 4

Because $f_i = \frac{x_i}{x}$ and $m_i = \frac{y_i}{y}$, from the MSE condition (10), we have

$$\frac{y_i}{y} \sum_j f_j P_j = \frac{x_i}{x} P_i.$$

Thus $\frac{x_i}{y_i} = \Gamma \frac{\overline{P}_{pr}}{P_i}$ (then $\frac{y_i}{x_i} = \frac{P_i}{\Gamma \overline{P}_{pr}}$).

When we substitute this coefficient to payoff functions of males, which take the form (in terms of auxiliary symbols $\Gamma$ and $\overline{P}_{pr}$) $W_m = W_{mm} + \frac{x_i}{y_i} W_{fm} = 0.5 \Gamma \overline{P}_{pr} k + \frac{x_i}{y_i} 0.5 P_i k$, then coefficient $\frac{x_i}{y_i} 0.5 P_i k$, which describes the per capita normalized number of new male individuals produced by female carriers of strategy $P_i$, will be equal to $0.5 \Gamma \overline{P}_{pr} k$. In effect,

$W_m = \Gamma \overline{P}_{pr} k$, which is the proof of point (i).

The analogous operation for the female payoff function takes the form:

$$W_f = W_{ff} + \frac{y_i}{x_i} W_{mf} = 0.5(1 - P_i)k + \frac{y_i}{x_i} 0.5(1 - \overline{P}_{pr})\Gamma k \ .$$

By substituting the transformed MSE condition $\frac{y_i}{x_i} = \frac{P_i}{\Gamma \overline{P}_{pr}}$ in this function, we obtain $\frac{y_i}{x_i} W_{mf} = 0.5 P_i \frac{1 - \overline{P}_{pr}}{\overline{P}_{pr}} k$ . After the substitution of the condition $\overline{P}_{pr} = P$ , we obtain

$\frac{y_i}{x_i} W_{mf} = 0.5 P_i \frac{1 - P}{P} k = 0.5 \Gamma P_i k$ . In effect, the female payoff function will have the form

$\frac{1}{2} k \left( (1 - P_i) + P_i \Gamma \right)$, which is the proof of point (ii).

Since $f_i = \frac{x_i}{x}$, $m_i = \frac{y_i}{y}$ and $\Gamma = \frac{x}{y}$, the MSE condition $f_i P_i = m_i \sum_j f_j P_j$ can be described as:

$$x_i P_i k = y_i \Gamma \overline{P}_{pr} k$$

The left side of this formula describes the number of male individuals produced by all female carriers of gene $P_i$, and the right side describes the number of male individuals produced by female partners of male carriers of the gene, which is the proof of point (iii).

## Appendix 5

## Proof of Lemma 5

Recall that the subpopulation $\Lambda$ is described by the vector of indices $d^\Lambda = [d_1^\Lambda, ..., d_{u^\Lambda}^\Lambda]$ . Thus $x_{d_i^\Lambda}$

$\left( y_{d_i^\Lambda} \right)$ is the number of females (males) with strategy $d_i^\Lambda$ and $f_{d_i^\Lambda} = \frac{x_{d_i^\Lambda}}{x} \ \left( m_{d_i^\Lambda} = \frac{y_{d_i^\Lambda}}{y} \right)$ is the respective frequency in the female (male) subpopulation. The average female strategy in the subpopulation $\Lambda$ is $\overline{P}_{pr}^\Lambda = \sum_i \frac{x_{d_i^\Lambda}}{x^\Lambda} P_{d_i^\Lambda}$ , where $x^\Lambda = \sum_i x_{d_i^\Lambda}$ , denotes the number of females in subpopulation $\Lambda$ and $y^\Lambda = \sum_i y_{d_i^\Lambda}$ represents the number of males. When we sum up the MSE conditions (10) of all strategies in subpopulation $\Lambda$, we obtain:

$$\sum_i m_{d_i^\Lambda} \sum_j f_j P_j = \sum_i f_{d_i^\Lambda} P_{d_i^\Lambda} .$$

In terms of exact numbers instead of related frequencies, this formula takes the form:

$$\frac{\sum_i y_{d_i^\Lambda}}{y} \sum_j f_j P_j = \frac{\sum_i x_{d_i^\Lambda} P_{d_i^\Lambda}}{x} .$$

Then

$$\frac{y^\Lambda}{y} \sum_j f_j P_j = \frac{x^\Lambda \sum_i \dfrac{x_{d_i^\Lambda}}{x^\Lambda} P_{d_i^\Lambda}}{x} .$$

When we multiply this condition by $\dfrac{x}{y^\Lambda}$, we obtain:

$$\frac{x}{y} \overline{P}_{pr} = \frac{x^\Lambda}{y^\Lambda} \overline{P}_{pr}^\Lambda$$

that is,

$$\Gamma \overline{P}_{pr} = \Gamma^\Lambda \overline{P}_{pr}^\Lambda ,$$

constituting the end of the proof.

$\boxed{\cdot}$

# Appendix 6

## Proof of Lemma 6

The gene fitness function has the form:

$$W_g(P_i, G, M) = \frac{k}{2}\big((1 - M_i) + \Gamma M_i\big).$$

According to Lemma 1, the sex ratios in carrier subpopulations are affected by male

subpopulation equilibrium and according to the results from [6] rapidly converge to:

$$\overline{\overline{M}}_i = \frac{P_i}{\Gamma \overline{P}_{pr} + P_i}.$$

When we substitute $\overline{\overline{M}}_i$ to $W_g(P_i, G, M)$, we obtain:

$$W_g(P_i, G, M) = \frac{k}{2}\left(1 - \frac{P_i}{\Gamma \overline{P}_{pr} + P_i} + \frac{\Gamma P_i}{\Gamma \overline{P}_{pr} + P_i}\right) = \frac{k}{2}\left(\frac{\Gamma \overline{P}_{pr} + P_i - P_i + \Gamma P_i}{\Gamma \overline{P}_{pr} + P_i}\right) =$$

$$= \frac{k}{2}\left(\frac{\Gamma(\overline{P}_{pr} + P_i)}{\Gamma \overline{P}_{pr} + P_i}\right) = \frac{k}{2}\left(\frac{\overline{P}_{pr} + P_i}{\overline{P}_{pr} + \frac{P_i}{\Gamma}}\right).$$

Thus the gene fitness function will then have the form:

$$W_g(P_i, G, M) = \frac{k}{2}\left(\frac{\overline{P}_{pr} + P_i}{\overline{P}_{pr} + \frac{P_i}{\Gamma}}\right).$$

In effect:

when $\frac{1}{\Gamma} < 1$, then a higher $P_i$ will have higher fitness (maximized by $P_i = 1$)

when $\frac{1}{\Gamma} > 1$, then a lower $P_i$ will have higher fitness (maximized by $P_i = 0$).

Because $\Gamma = \frac{1 - P}{P}$, equilibrium point $\Gamma = 1$ is equivalent to $P = 0.5$ and above rule has the following form:

when $P < 0.5$, then a higher $P_i$ will have higher fitness (maximized by $P_i = 1$)

when $P > 0.5$, then a lower $P_i$ will have higher fitness (maximized by $P_i = 0$).

This means that it is profitable to produce the sex which is currently in the minority. This condition is fully compatible with the classical theory, thus ending the proof.

## List of important symbols:

$y$ - number of males

$x$ - number of females

$u$ - number of individual strategies

$f_i = \dfrac{x_i}{x}$ - frequency of females with strategy $P_i$

$m_i = \dfrac{y_i}{y}$ frequency of males with strategy $P_i$

$f = [f_1, ..., f_u]$ - vector of the state of the female subpopulation

$m = [m_1, ..., m_u]$ vector of the state of the male subpopulation

$P = \dfrac{y}{y+x}$ - frequency of males in the population

$\Gamma = \dfrac{x}{y} = \dfrac{1-P}{P}$ - number of females per single male individual (auxiliary parameter)

$\overline{P}_{pr} = \sum_j f_j P_j$ - mean strategy in the female subpopulation

$G_i = P m_i + (1-P) f_i$ - frequency of a gene encoding $i-$ th strategy

$G = [G_1, ..., G_u]$ - vector of a state of a gene pool

$M_i = \dfrac{P m_i}{P m_i + (1-P) f_i}$ - fraction of males in the subpopulation of carriers of the $i-$ th strategy

$W_m (P_i, P, f, m) = k \dfrac{1-P}{2P} \left( \sum_j f_j P_j + \dfrac{f_i}{m_i} P_i \right)$ - male payoff function

$W_f (P_i, P, f, m) = \dfrac{k}{2} \left( (1-P_i) + \dfrac{m_i}{f_i} \left( 1 - \sum_j f_j P_j \right) \right)$ - female payoff function

$W_g (P_i, G, M) = \dfrac{k}{2} \left( \Gamma M_i + (1 - M_i) \right)$ - fitness function of a gene encoding strategy $P_i$

Figure legends:

Fig .1 Comparison of static fitness measures generated by the new model and used in classical theory. a) Plot of the classical Dusing-Fisher fitness measure denoted by $W$. b) Plot of a fitness measure generated by the new model denoted by $W_g$ (when the MSE condition is satisfied). c) Plot of differences between fitness measures, $W - W_g$. $P_i$ defines the individual strategy and $\overline{P}_{pr}$ defines the primary sex ratio.

Fig. 2 Sections of surfaces from Figure 1. The classical fitness measure is a linear function of parameter $P$, whereas the new fitness measure is nonlinear. a) Plot of the classical Dusing-Fisher

fitness measure denoted by $W$. b) Plot of a fitness measure generated by the new model denoted by $W_g$ (when the MSE condition is satisfied) c) Plot of differences between both fitness measures $W - W_g$.

Fig. 3 The classic DFSM fitness measure is fully compatible with predictions of the new model only in the case of a single unique mutant female. In the next generation, all carriers of a mutant gene are her offspring and the sex ratio in the population of her descendants carrying the mutant gene is equal to the value of expression of her strategy (see Figure a – a female with strategy 3/7). In further generations, there are a number of male carriers of a mutant gene (see Fig b), whose number depends on strategies represented by their partners. The offspring of these males are disregarded by classical theory, whereas the new model takes them into consideration. Bias of the classical fitness measure caused by the influence of males on a gene's reproductive success may lead to differences in the predictions of both theories.

Fig. 4 Mechanistic explanation of the male subpopulation equilibrium. Partners of every male produce the same number of offspring carrying his gene ($\frac{k}{2}\Gamma \overline{P}_{pr}$), so the differences in payoff are the effects of differences in the activity of female carriers of the same gene. The influence of female carriers is determined by the factor $\frac{x_i}{y_i}$ (the number of female carriers per single male carrier). If this factor has a high value (more females per single male), then male carriers of this gene will have a larger growth rate (this situation is presented in panel a). However, when the number of male carriers $y_i$ increases, then factor $\frac{x_i}{y_i}$ will decrease, which will cause a decrease in the influence of female carriers of the gene on the payoff of the average male carrier of the same gene (panel b). This process will lead to an equilibrium state in which the influence of female carriers on all male carriers is the same.

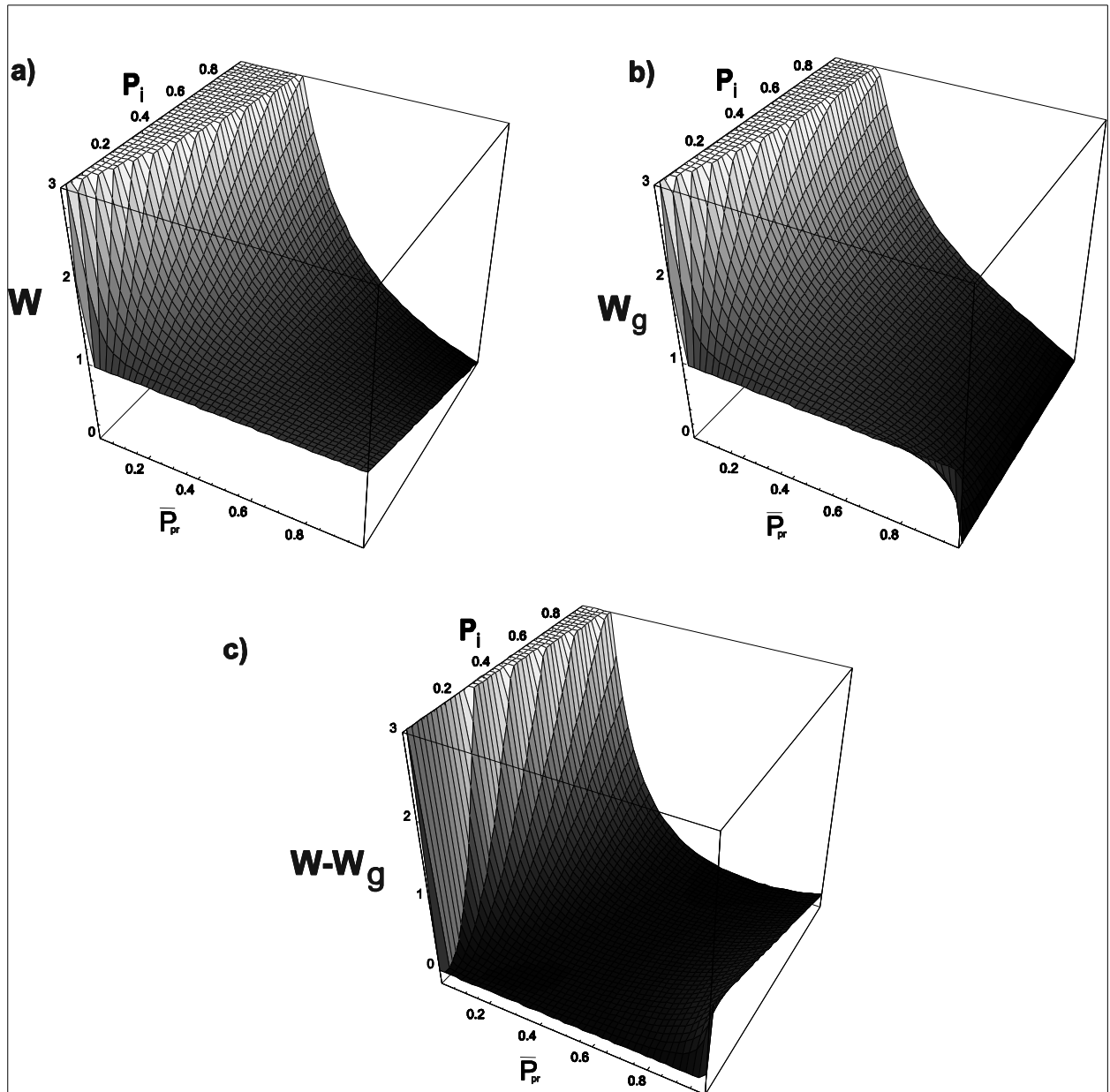

Fig. 1

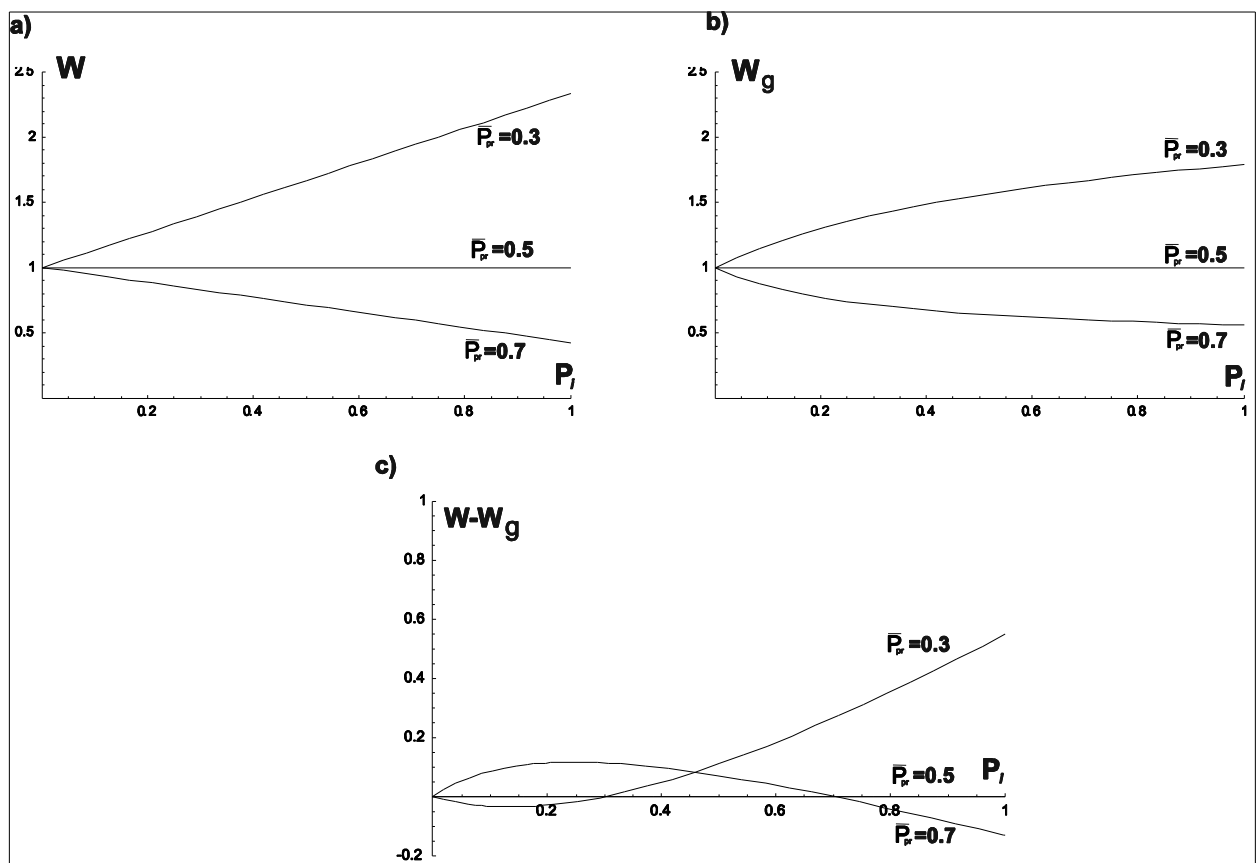

Fig. 2

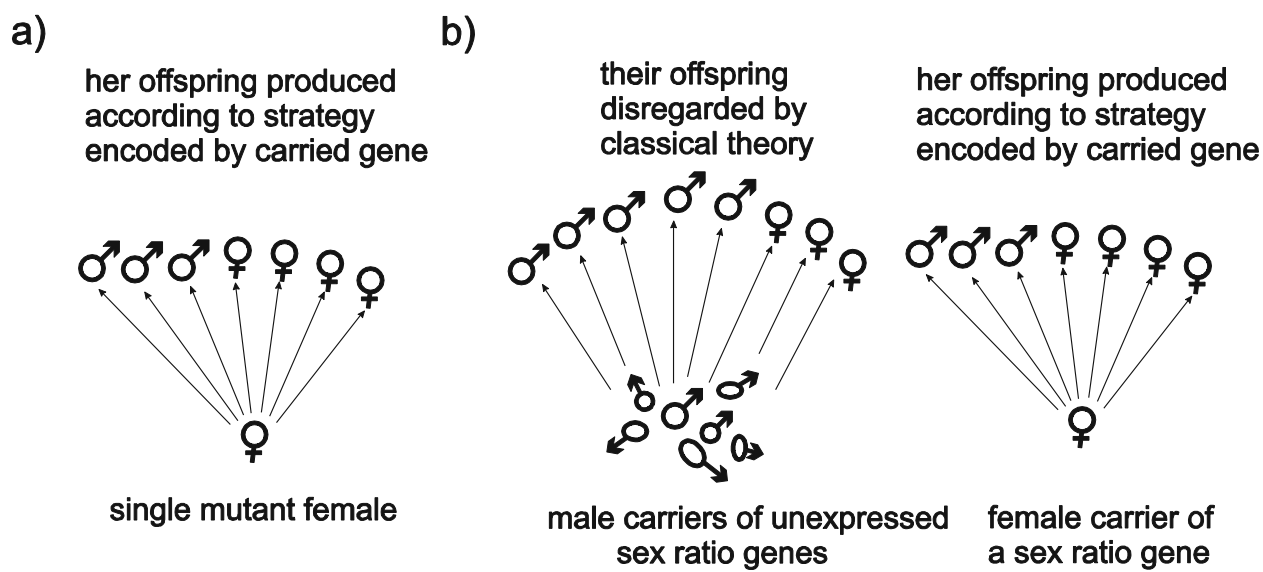

Fig. 3

a)

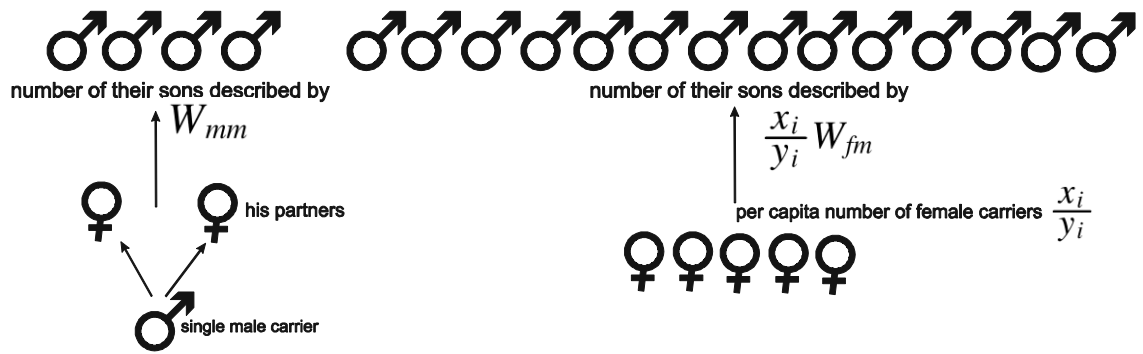

b)

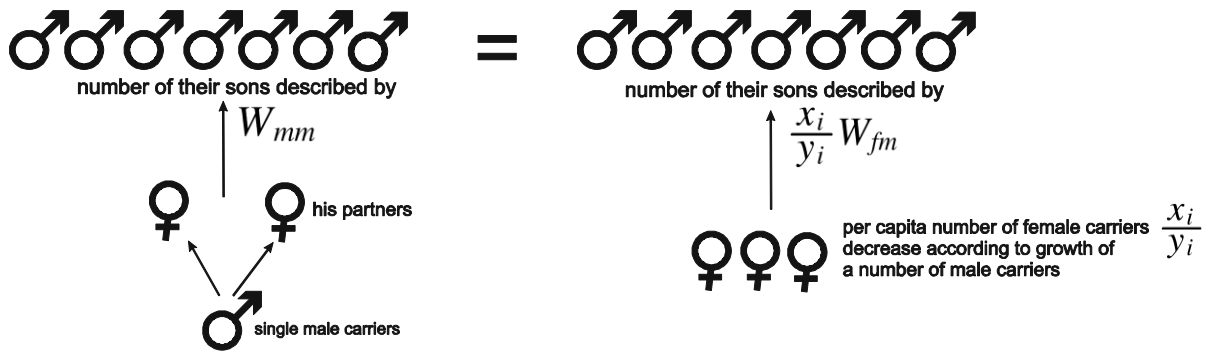

Fig. 4